\documentclass[12pt,a4paper,superscriptaddress,preprint,showpacs]{revtex4}
\usepackage{graphicx}
\usepackage{amsmath,amsfonts} 

\begin{document}

\newcommand{\EQ}{Eq.~}
\newcommand{\EQS}{Eqs.~}
\newcommand{\FIG}{Fig.~}
\newcommand{\FIGS}{Figs.~}
\newcommand{\SEC}{Sec.~}
\newcommand{\SECS}{Secs.~}


\title{Networks with Dispersed Degrees Save
Stable Coexistence of Species in Cyclic Competition}
\author{Naoki Masuda}
\affiliation{
Amari Research Unit, RIKEN Brain Science Institute,
2-1, Hirosawa, Wako, Saitama 351-0198, Japan}
%
\author{Norio Konno}
\affiliation{
Faculty of Engineering, Yokohama National University,
79-5, Tokiwadai, Hodogaya, Yokohama 240-8501, Japan}
%
\date{Received 28 July 2006}

\begin{abstract}
Coexistence of individuals with 
different species or phenotypes is
often found in nature in spite of competition between them. Stable
coexistence of multiple types of individuals
have implications for maintenance of ecological biodiversity
and emergence of altruism in society, to name a few.
Various mechanisms of coexistence 
including spatial structure of populations, heterogeneous
individuals, and heterogeneous environments, have been proposed.
In reality, individuals disperse and interact on
complex networks.  We examine
how heterogeneous degree distributions of networks
influence coexistence, focusing on models of cyclically competing species.
We show analytically and numerically that 
heterogeneity in degree distributions promotes
stable coexistence.
\end{abstract}


\maketitle


\newpage

\section{Introduction}\label{sec:introduction}

How to maintain or prevent coexistence of competing multiple types of
individuals is a key issue in various areas.
For example, coexistence of multiple species in ecological habitats
implies stable biodiversity realized in nature 
\cite{Connell78,Chesson00}. Coexistence of multiple types of players
in evolutionary games implies survival of altruistic players in the
sea of selfish players \cite{Axelrod84}. Coexistence of disease-free
and infected individuals implies an endemic state that should
be suppressed usually \cite{Andersonbook}.

Mechanisms of coexistence have been a
central theoretical question because complexity of a population state
(i.e. coexistence) and stability are often contradicting requirements
\cite{May72,Tilman94,Hofbauerbook}. Coexistence in population dynamics
has been explained by, for example,
nonequilibrium-state interpretation, habitat subdivision,
heterogeneity in species such as heterogeneous dispersal speeds, and
heterogeneity in environments
\cite{Connell78,Chesson00,Tilman94,
Murdoch92}. 
Spatial structure such as the square lattice
also limits diffusion and promotes coexistence.  In this case, each
species is clustered in different regions of the lattice
\cite{Hassell91,Tainaka93,Durrett98TPB,Frean01}.
However, real-world interaction quite often occurs on contact networks
of individuals that are more complex than the square lattice.
Most real networks have the small-world and scale-free properties
(e.g. Refs. \cite{Barabasi99,NewmanSIAM}).
The small-world property is equivalent to the
combination of small average distance between vertices and large
clustering, or abundance of densely connected small subgraphs such as
triangles.  A scale-free network is defined by a degree distribution,
or the distribution of the number of contacts (edges) that each vertex
has, which follows a power law, $p_k\propto k^{-\gamma}$. Here $p_k$
is the probability that a vertex has degree $k$.
The scale-free property may be too idealistic to describe
contact networks underlying real population dynamics.  Even so, it
seems likely that different patches or individuals 
are endowed with different connectivity to others.

In terms of networks, some known mechanisms of coexistence benefit from
the regular lattices, the one-dimensional continuous line, the
two-dimensional continuous plane, and the complete graph (mean field
situation), in which all the vertices are considered to share the same
degree (see the papers cited above and references therein).
In random graphs, which are sometimes used in this context
\cite{May72}, the vertex degree obeys the Poisson
distribution. However, the real degree distribution may be even
broader.

Here we investigate how possibility of coexistence is affected by
heterogeneous degree distributions of contact networks, not by
heterogeneous environments other than network-based ones or
heterogeneous individuals. Among various
competitive relationships among different phenotypes, we focus on
cyclic competition of three species, which is a minimal case.

Cyclic competition 
is actually abundant in nature. For example, tropical marine
ecosystems \cite{Buss80} and vertebrate communities in high-arctic
areas \cite{Gilg03}
include cyclic dominance relations composed of a couple of
organisms
(also see Ref. \cite{Hofbauerbook}).
Real microbial communities of {\it
Escherichia coli} \cite{Kerr02}
%
%
and color polymorphisms of
natural lizards \cite{Sinervo96} also have cyclically dominating three
phenotypes and show alternating wax-and-wane population
densities.
%
%
In evolutionary games,
the public-good game with volunteering, namely, the choice of not
joining the game, results in cyclic competition
%
\cite{Hauert02_Szabo02_Semmann03}.
The susceptible-infected-recovered-susceptible model of epidemiology
and models with additional types of states also include 
cyclic competition \cite{Andersonbook,MasudaCPS}.
We focus on two specific predator-prey models of such cyclic interaction,
that is, the standard
rock-scissors-paper (RSP) model \cite{Hofbauerbook} and
the May-Leonard (ML) model \cite{May75}.  These models have neutrally
stable or unstable coexistence solutions in well-mixed populations.
Therefore, in a finite population, population dynamics are eventually
trapped by an absorbing state corresponding to the dominance of one
species \cite{Hofbauerbook,Taylor78}.  We show that heterogeneous degree
distributions stabilize coexistence of multiple types of individuals
placed on networks.

\section{Rock-Scissors-Paper Dynamics on Networks}

\subsection{Model}

As a minimal model of cyclic competition, we consider the 
standard RSP dynamics on networks with heterogeneous
contact rates.  There are three species, which we call states,
represented by rock,
scissors, and paper; rock beats scissors, scissors beat paper, and
paper beats rock.  Each vertex takes state 0, 1, or 2.  A pair of
vertices may be connected by an edge.
The degree $k$ of a vertex is the number of edges, or the number of
contacts with other vertices. State 1
outcompetes state 0 by invading onto each
neighboring state with state 0 at a rate of $\lambda$. In other words,
a vertex with state 0 changes its state to 1 at a rate of $\lambda
n_1$, where $n_i$ is the number of vertices with state $i$ in the
neighborhood.  Similarly, 1 (2) turns into 2 (0) at a rate of $\mu
n_2$ ($n_0$). In an ecological context,
we are considering the limit that dispersion rates ($=$ 1,
$\lambda$, $\mu$) are much larger than the natural death
rate.
We consider the influence of death rates later with
the ML model.


For a perfectly mixed population, 
the mean field theory tells that there are an ensemble of neutrally
stable limit cycles surrounding a neutrally stable equilibrium
corresponding to coexistence of the three states.  Therefore, the
coexistence solution is practically unstable in finite populations.
The RSP dynamics with spatial structure, such as
the square lattice, accommodate many states each of which is clustered
in different loci
\cite{Tainaka93,Frean01,Durrett98TPB}.
Here we are interested in a network mechanism that may enable
stable coexistence.


\subsection{Equilibrium}

With dispersed degrees, vertices with
different degrees obey different state-transition dynamics.
Let us denote by $\rho_{i,k}$ the probability that a vertex with
degree $k$ takes state $i$ ($=0, 1, 2$). The probability that a vertex
adjacent to an arbitrary vertex takes state $i$ is denoted by
$\Theta_i$. This probability does not generally agree with
$\rho_{i,k}$ or its average over all the vertices.  This is because a
vertex with more edges is more likely to be selected as a neighbor. In
fact, a neighbor has degree $k$ with probability $k
p_k/\left<k\right>$, where $p_k$ is the probability that a vertex has 
degree $k$ and 
$\left<k\right>=\sum k p_k$ is the mean degree giving 
normalization. Therefore,
%
%
$\Theta_i = \sum_k k p_k \rho_{i,k} / \left< k\right>$
\cite{Hethcote84_Pastor01PRL}.
%
Because each vertex is occupied by one of the three species, namely,
$\rho_{0,k} = 1 - \rho_{1,k} - \rho_{2,k}$ and $\Theta_0 = 1 - 
\Theta_1 - \Theta_2$, it suffices to consider the density of state 1
and that of state 2. Noting that the expected number of state-$i$ 
neighbors of a vertex with degree $k$ is equal to
$k\Theta_i$, we derive
\begin{eqnarray}
\dot{\rho}_{1,k} &=& \lambda (1-\rho_{1,k}-\rho_{2,k}) k
\Theta_1 - \mu \rho_{1,k} k \Theta_2,
\label{eq:rock_aa}\\
\dot{\rho}_{2,k} &=& \mu \rho_{1,k} k
\Theta_2 - \rho_{2,k} k (1-\Theta_1-\Theta_2).
\label{eq:rock_a}
\end{eqnarray}
For example, 
the first term in \EQ(\ref{eq:rock_aa}) corresponds to the invasion
of state 1 onto vertices with state 0.
In the equilibrium, we have
\begin{equation}
\left(\begin{array}{c}
\rho^*_{1,k}\\ \rho^*_{2,k}
\end{array}\right)
= \frac{\lambda \Theta^*_1}{(\lambda\Theta^*_1+\mu\Theta^*_2)
(1-\Theta^*_1-\Theta^*_2)+\lambda\mu\Theta^*_1\Theta^*_2}
\left(\begin{array}{c}
1-\Theta^*_1-\Theta^*_2\\
\mu\Theta^*_2
\end{array}\right).
\label{eq:rock_b}
\end{equation}
The coexistence solution to \EQ(\ref{eq:rock_b})
is given by
\begin{equation}
\left(\begin{array}{c}
\Theta^*_1\\ \Theta^*_2
\end{array}\right)
= \left(\begin{array}{c}
\rho^*_{1,k}\\ \rho^*_{2,k}
\end{array}\right)
= \frac{1}{\lambda+\mu+1}
\left(\begin{array}{c}
1\\ \lambda
\end{array} \right),
\label{eq:rock_c}
\end{equation}
for any $k$. The degree distribution 
does not affect the equilibrium
population densities \cite{MasudaCPS}. 

\subsection{Stability of coexistence equilibrium}

When $p_k = \delta_{k,\left<k\right>}$, each vertex
has the same degree equal to the mean $\left<k\right>$.
This case corresponds to
well-mixed populations. Then the coexistence is neutrally stable
(e.g. Refs. \cite{Frean01,May75}), which underlies experimental and natural
ecosystems showing oscillatory population dynamics
\cite{Gilg03,Kerr02,Sinervo96}.
%
%
Equation~(\ref{eq:rock_a})
indicates that the oscillation period is proportional to
$1/\left<k\right>$.

However, the stability of the coexistence and
realized dynamics are considerably influenced by networks.
To see this, let us consider a two-point degree distribution given by
$p_k = p\delta_{k,k_1} + (1-p)\delta_{k,k_2}$.
On average, a total of $np$ vertices have degree $k_1$ and $n(1-p)$
vertices have degree $k_2$.
Equations~(\ref{eq:rock_aa}) and (\ref{eq:rock_a})
for a network with the two-point degree
distribution define a
four-dimensional dynamical system.  We set $\lambda=\mu=1$ for
simplicity, although generalization to other
$\lambda$ and $\mu$ is straightforward. The characteristic equation
evaluated at the coexistence equilibrium [\EQ(\ref{eq:rock_c})] is
represented by
\begin{equation}
x^4 + \frac{3 k_1 k_2}{\left<k\right>}x^3 +
3\left( \frac{k_1 k_2\left(k_1 + k_2 - \left<k\right>\right)}{\left<k\right>} 
+ \frac{\left<k^2\right>^2}{\left<k\right>^2} \right) x^2
+ 9\frac{\left<k^2\right>}{\left<k\right>}k_1 k_2 x
+ 9 k_1^2 k_2^2 = 0,
\label{eq:ch_rsp}
\end{equation}
where $\left<k\right> = \sum k p_k 
=p k_1 + (1-p) k_2$ and $\left<k^2\right> =
\sum k^2 p_k = p k_1^2 + (1-p) k_2^2$.
When $k_1 = k_2 = k$, we turn back to the ordinary mean field case with
neutrally stable oscillations:
$x=\sqrt{3k}i$, $(-3\pm
\sqrt{3})k/2$. More generally, the Routh-Hurwitz criteria for
\EQ(\ref{eq:ch_rsp}) is
\begin{eqnarray}
|H_1| &=& 1,\nonumber\\
|H_2| &=& p k_2^2\left(\left<k\right>-k_1/2\right)^2
+ (1-p) k_1^2\left(\left<k\right>-k_2/2 \right)^2 + 3k_1^2 k_2^2/4 > 0, 
\nonumber\\
|H_3| &=&  81 k_1^2 k_2^2
p(1-p)(k_2-k_1)^2\left(\left<k^2\right>^2 + 2 k_1 k_2 \left<k\right>^2
\right) \big/\left<k\right>^4,\nonumber\\
|H_4| &=& 9 k_1^2 k_2^2 |H_3|,
\end{eqnarray}
where $|H_i|$ is the $i$th principal minor.
The coexistence solution is stable when $|H_3|, |H_4| > 0$, that is,
$k_1\neq k_2$ and $p\neq 0, 1$. Dispersed contact rates
stabilize coexistence.

\subsection{Numerical results}

We resort to numerical simulations to examine more general networks
and to be more quantitative about the effects of degree dispersion.
We compare different types of networks with $n=5000$ vertices and
$\left<k\right> = 10$.  The regular (R) random graph corresponding to
the ordinary mean field case is generated by the configuration model
\cite{NewmanSIAM} with $p_k = \delta_{k,\left<k\right>}$. This is a
type of random graph in which every vertex has the same degree
$\left<k\right>$.  We also use the Erd\"{o}s-R\'{e}nyi (ER) random
graph, which has the Poisson degree distribution $p_k =
e^{-\left<k\right>}\left<k\right>^k / k!$, and the Barab\'{a}si-Albert
(BA) scale-free network with the parameter 
$m\cong \left<k\right>/2 = 5$, which
yields $p_k \propto k^{-3}$ ($k\ge m$) and $p_k=0$ ($k<m$)
\cite{Barabasi99}.

Typical population dynamics for these networks are shown in
\FIGS\ref{fig:rsp}(a)--\ref{fig:rsp}(c).
On the R random graph, the coexistence
solution is neutrally stable in theory.  Combined with a finite-size
effect, the amplitude of the dynamical population density becomes
progressively large in an oscillatory fashion.  Eventually, one state
dominates the whole network in an early stage [\FIG\ref{fig:rsp}(a)].
On the ER random graph [\FIG\ref{fig:rsp}(b)] and the BA model
[\FIG\ref{fig:rsp}(c)], coexistence occurs owing to the distributed
$k$.  The fluctuations in the population density is smaller on the BA
model than on the ER graph, because the BA model has a broader 
degree distribution.

To be more systematic, we compare population density fluctuation in the
coexistence equilibrium.  The fluctuation 
is measured by the standard deviation of
the time series $\rho_i$ [see \FIG\ref{fig:rsp}(b) and
\ref{fig:rsp}(c)] after
transient, averaged over $i=0, 1$, and $2$. Larger fluctuation 
means more unstable coexistence, and we examine how the size of the 
fluctuation depends on the amount of degree dispersion.
In addition to the
networks examined above, we use two types of networks that can
create a range of degree dispersion. The first is the network with
the two-point degree distribution.  The standard deviation of the degree
$\sqrt{\left<k^2\right>-\left<k\right>^2} =
\sqrt{p(1-p)}|k_1-k_2|$. By varying $k_1/k_2$ with $p=0.9$ and
$\left<k\right>=10$ fixed, we can systematically create networks with
a variety of $\sqrt{\left<k^2\right>-\left<k\right>^2}$. The second
is the network that has
Gaussian $p_k$ with mean $\left<k\right>$, whose
$\sqrt{\left<k^2\right>-\left<k\right>}$ can be also modulated.
The results are summarized
in \FIG\ref{fig:rsp}(d) for four types of networks (ER, BA, two-point,
and Gaussian), excluding the R random graph because it does not sustain
coexistence.  Regardless of the network type, more
dispersed degree distributions generally lead to more stable coexistence.

In the mean field case,
a smaller network with a stronger finite-size effect tends to
drive the population dynamics to the absorbing equilibrium where only one
state survives \cite{Taylor78}.  The network effect on stability of
coexistence is more manifested in this regime.  In
\FIG~\ref{fig:rsp_surv}, we show survival probabilities for some
networks with $n=200$, where the survival is defined by existence of
all the three states. Figure~\ref{fig:rsp_surv} is consistent with
\FIG\ref{fig:rsp}(d); coexistence is sustained for a
longer period on networks with larger degree dispersion.

\section{May-Leonard Dynamics on Networks}

\subsection{Model and equilibrium}

Since neutrally stable oscillations of the RSP model may be singular
phenomena, we analyze another competition model proposed by May and
Leonard \cite{May75}.  The ML model represents dynamics of cyclically
competing three species with natural death. Because of the natural death, 
vertices can take the vacant state. In a well-mixed
population,
the coexistence equilibrium and the periodic
oscillation are both unstable.  A trajectory of the population density
approaches heteroclinic orbits on which at least one of the three
species is extinct. Theoretically, one species is
transiently and alternatively dominant with ever increasing
periods in an infinite population.
Practically, one species eventually wins due to the
finite-size effect.

As an interacting particle system, the ML model is
a four-state system, with state 0 representing the vacant
site and 1, 2, and 3 representing cyclically dominating states
\cite{Durrett98TPB}.  The ML
dynamics with heterogeneous contact rates are written as
\begin{eqnarray}
\dot{\rho}_{1,k} &=& \rho_{0,k}k\Theta_1 -
(\alpha-1)\rho_{2,k}k\Theta_1 - (\beta-1)\rho_{1,k}k\Theta_3,
\nonumber\\
\dot{\rho}_{2,k} &=& \rho_{0,k}k\Theta_2 -
(\alpha-1)\rho_{3,k}k\Theta_2 - (\beta-1)\rho_{2,k}k\Theta_1,
\label{eq:may}
\\
\dot{\rho}_{3,k} &=& \rho_{0,k}k\Theta_3 - 
(\alpha-1)\rho_{1,k}k\Theta_3 - (\beta-1)\rho_{3,k}k\Theta_2,\nonumber
\end{eqnarray}
where
$\rho_{0,k} = 1 - \rho_{1,k} - \rho_{2,k} -
\rho_{3,k}$. The first term in each equation indicates the rate at which
a vacant site becomes colonized by state $i$ ($1\le i\le 3$).
Supposing that $\alpha<1$ and $\beta>1$, 
the second term and the third term of each equation
represent the population increase and decrease due to the cyclic
competition, respectively. For example, in the first equation,
state 1 outcompetes 2, whereas 1 is outcompeted by 3.
Equation~(\ref{eq:may}) has a coexistence equilibrium given by
\begin{equation}
\rho^*_{i,k} = \Theta^*_i = (\alpha+\beta+1)^{-1}\quad (i=1,2,3)
\end{equation}
for all $k$.  
With
$p_k=\delta_{k,\left<k\right>}$,
the eigenvalues of the Jacobian matrix evaluated
at ($\rho^*_{1,k}$, $\rho^*_{2,k}$, $\rho^*_{3,k}$), disregarding a
prefactor $(\alpha+\beta+1)^{-1}$, are $x = -(1 + \alpha + \beta), -1
+ (\alpha+\beta)/2 \pm \sqrt{3}(\alpha-\beta)i/2$ \cite{May75}.
When $\alpha + \beta = 2$, the ML model is 
essentially the same as
the RSP model, and
the coexistence is neutrally stable [$x=-3, \pm
\sqrt{3}(\alpha-\beta)i/2$].  
When
$\alpha + \beta > 2$ and $\alpha<1$ (or $\beta<1$), the coexistence
equilibrium is an unstable spiral, and the trajectory
tends to a homoclinic orbit.

\subsection{Stability of coexistence equilibrium}

To investigate the network effect, we again consider the two-point
degree distribution $p_k = p\delta_{k,k_1} + (1-p)\delta_{k,k_2}$.
The Jacobian matrix at $(\rho^*_{1,k_1}, \rho^*_{1,k_2},
\rho^*_{2,k_1}, \rho^*_{2,k_2}, \rho^*_{3,k_1}, \rho^*_{3,k_2})^t$,
where $t$ denotes the transpose, is a block circulant matrix.
Accordingly, the eigenmode
has the form $(v_1, v_2, \rho v_1, \rho v_2, \rho^2 v_1, \rho^2
v_2)^t$, $v_1, v_2\in {\mathbf C}$, where $\rho$ is any solution to
$\rho^3 = 1$.  The corresponding characteristic equation is reduced to
\begin{eqnarray}
x^2 &+& \left( (\beta+\rho\alpha+\rho^2)(k_1+k_2)
- (1-\rho^2)(\beta-1)\frac{\left<k^2\right>}{\left<k\right>}
\right)x\nonumber\\
&+& (\alpha^2+\beta^2-\alpha\beta-\alpha-\beta+1)\rho^2 k_1 k_2= 0.
\label{eq:eig_may}
\end{eqnarray}
For $\rho=1$, \EQ(\ref{eq:eig_may}) is a real equation, and two
eigenvalues have the same real part that is equal to
$-(\alpha+\beta+1)(k_1+k_2)/2 < 0$. Because an eigenvalue of
\EQ(\ref{eq:eig_may}) for $\rho = \exp(2\pi i/3)$ is conjugate of one
for $\rho = \exp(4\pi i/3)$, it suffices to set $\rho = \exp(2\pi
i/3)$. The larger real part of the
solution to \EQ(\ref{eq:eig_may}) is plotted in
\FIG\ref{fig:eigs_may} for various $p$ and $0<k_1/k_2\le 1$.
The coexistence
equilibrium is stabilized with negative real parts of the eigenvalues.
This occurs when $k_1/k_2\cong 0.1$ and $p=0.9, 0.95$.
In this situation, a small 
number $[=n(1-p)]$ of hubs have
a large degree $(=k_2)$ in comparison with
most vertices with degree $k_1$. This is reminiscent of
long-tail $p_k$ typical of the scale-free networks.
Excess heterogeneity ($k_1/k_2 <
0.05$) destroys coexistence. In this situation, a majority of
vertices with degree $k_1$ are effectively
isolated, and the network is close to the mean field case, or 
the R random graph with $k=k_2$.
When $p<0.5$, the heterogeneity
does not cause stability irrespective of $k_1/k_2$.
This is because the contribution of the smaller subpopulation (proportion
$p$) with the smaller degree $k=k_1 (<k_2)$
to dynamics is marginal,
which again results in effectively homogeneous networks
with $k=k_2$.

\subsection{Numerical results}

For numerical simulations, we note
that, in \EQ(\ref{eq:may}), 
a vacant site (state 0) is replaced by
state $i$ ($1\le i\le 3$) at a rate of $n_i$.
A vertex in
state 1 (2, 3) kills a neighboring state 2 (3, 1) at a rate
of $\beta-1$. Then, the neighboring vertex is colonized by state 1 (2,
3) with probability $(1-\alpha)/(\beta-1)$ and turns empty
(state 0) otherwise \cite{Durrett98TPB}.

Dynamics for the R and ER random graphs with $n=5000$ and
$\left<k\right>=10$ are shown in \FIGS\ref{fig:may}(a) and 
\ref{fig:may}(b),
respectively.  Because the stability condition for the ML model is
more severe than that for the RSP model, one state shortly
overwhelms the others on the ER as well as R random graph. However,
the transients for the ER random graph, whose degrees are more
dispersed than the R random graph, are longer.  The BA model and the
networks with two-point $p_k$ with parameters realizing the stable
Jacobian matrix (but not the networks with Gaussian $p_k$) yield
coexistence.  Similar to the RSP dynamics, the amount of degree
dispersion is strongly correlated with
the amount of density fluctuation and stability of
coexistence, irrespective of the network type
[\FIG\ref{fig:may}(c)].

\section{Discussion}

\subsection{Summary of the results}

We have examined population dynamics with 
cyclic dominance relationships on networks. The steady population
density is independent of degree distributions of 
networks.  However, stability of coexistence
equilibria and dynamics depend considerably on networks.
Heterogeneity in degree distributions facilitates stable
coexistence of different phenotypes. As touched upon in
\SEC\ref{sec:introduction}, coexistence of competing species
is desirable in, for example,
ecological communities (biodiversity) and evolutionary games (survival
of altruistic players).

\subsection{Relations to synchronization of coupled oscillators}

The present mechanism of coexistence is related to synchronization of
coupled oscillators.
With spread degree distributions, each
$\rho_{i,k}$ evolves at a speed proportional to $k$, and $\rho_{i,1}$,
$\rho_{i,2}$, $\ldots$ are coupled by a sort of mean field feedback
$\Theta_i$.  Then, the population dynamics are analogous to those of
an ensemble of phase oscillators coupled by mean field interaction. In
theory, coupled phase oscillators become desynchronized when the intrinsic
frequency of the oscillators has a broad distribution relative to the
coupling strength \cite{Kuramotobook}.  Oscillators with heterogeneous
intrinsic frequencies correspond to vertices with heterogeneous $k$.
In terms of competition dynamics
on networks, asynchrony corresponds to stable coexistence of
species where
synchronous oscillations (large fluctuations in time) 
of the population density is
suppressed.  Desynchronization is known to suppress neutrally stable
or unstable oscillations in ecological models with
patchy populations, heterogeneous
birth rates, and weak aggregation \cite{Murdoch92}.
Heterogeneity in degree distributions
serves to stable coexistence via desynchronization even without
other kinds of heterogeneity. The correspondence between asynchrony
and coexistence may be exported to more general models,
particularly to ones showing oscillations
in well-mixed populations; oscillatory population densities
are reminiscent of cyclic competition.

\subsection{Difference from spatial mechanisms of coexistence}

The scenario to coexistence unraveled here is
distinct from those based on spatial structure, heterogeneous
environments, or heterogeneous individuals. In patchy
habitats with heterogeneous environments or small
diffusion \cite{Tilman94}, and in spatial structure
with limited diffusion
\cite{Hassell91,Tainaka93,Durrett98TPB,Frean01}, multiple
species can coexist by forming locally high densities of conspecifics
in different subspaces
\cite{Tainaka93,Durrett98TPB,Frean01}. 
This is the spatial mechanism of coexistence.
In networks with dispersed degrees, 
multiple species can
coexist on a network in a mixed manner without spatial segregation.

Real networks of contacts are equipped with the clustering property,
as is the case for the regular lattices and small-world networks, and
high clustering elicits the spatial mechanism of coexistence.  In
addition, many networks own broad degree
distributions represented by the scale-free distributions
\cite{Barabasi99,NewmanSIAM,Hethcote84_Pastor01PRL}.  The mechanism
proposed in this work can cooperate with
the spatial mechanism to promote stability of coexistence.

\subsection{Difference from contagion dynamics}

There are many possible rules for interacting particle systems.
In contagion processes,
such as the percolation, the susceptible-infected-recovered
model, and the contact process (susceptible-infected-susceptible
model), degree dispersion affects
dynamical aspects by, for example, accelerating disease propagation
in initial stages \cite{Barthelemy04PRL}. More fundamentally, however,
epidemic thresholds (critical
infection rates) are proportional to $\left<k \right>/\left<k^2\right>$.
Then, disease propagation on a global scale
is more likely to occur on networks with more
heterogeneous contact rates with small
$\left<k \right>/\left<k^2\right>$
than on networks with rather homogeneous degrees such as the 
regular lattices and the random graph
\cite{Hethcote84_Pastor01PRL,Andersonbook}.
%
%
In contagion dynamics, 
the network influences the stationary state in addition to the
dynamics, which contrasts to our results for the competition dynamics.

Generally speaking, the positions of 
equilibria move when we cannot neglect at
least one type of state-transition event whose occurrence rate is
independent of neighbors' states $n_i$ \cite{MasudaCPS}. Examples are
spontaneous recovery and mutation. By contrast, the
equilibria are 
invariant if all the state transitions are controlled by the neighbors'
states, as is the case for the RSP model, the ML model, and also the
voter model.  However, the degree distribution does influence the
stability of coexistence and hence the whole population dynamics.  Our
results are generalized to other population dynamics in which the
rates of spontaneous transitions can be ignored.


\begin{acknowledgements}
The authors thank Michio Kondoh for valuable comments.
This work is supported by Special
Postdoctoral Researchers Program of RIKEN.
\end{acknowledgements}

\newpage

\pagestyle{empty}
Figure captions

\bigskip

Figure 1: RSP dynamics on networks with $n=5000$
and $\left<k\right>=10$.  The
initial condition is given by the Bernoulli distribution with $\rho_0
= 0.7$ and $\rho_1 = \rho_2 = 0.15$, where $\rho_i$ is the proportion
of vertices that take state $i$.  The densities $\rho_0$ (thin lines),
$\rho_1$ (moderate lines), and $\rho_2$ (thick lines) are shown for
(a) the R random graph.  For (b) the ER random graph and
(c) the BA model of the same size, only $\rho_0$ is shown for clarity.
(d) Fluctuation of population density as a function of the standard
deviation of the vertex degree (ER, triangle; BA, horizontal line
[$\left(\left<k^2\right>-\left<k\right>^2\right)^{1/2} = 157.8$];
Gaussian $p_k$, crosses; two-point $p_k$, circles).
The variance of 
$\rho_i$ from time 150 through 300 averaged over
$i=0, 1$, and $2$ defines the density fluctuation.

\bigskip

\begin{figure}
\begin{center}
\includegraphics[height=3in,width=3in]{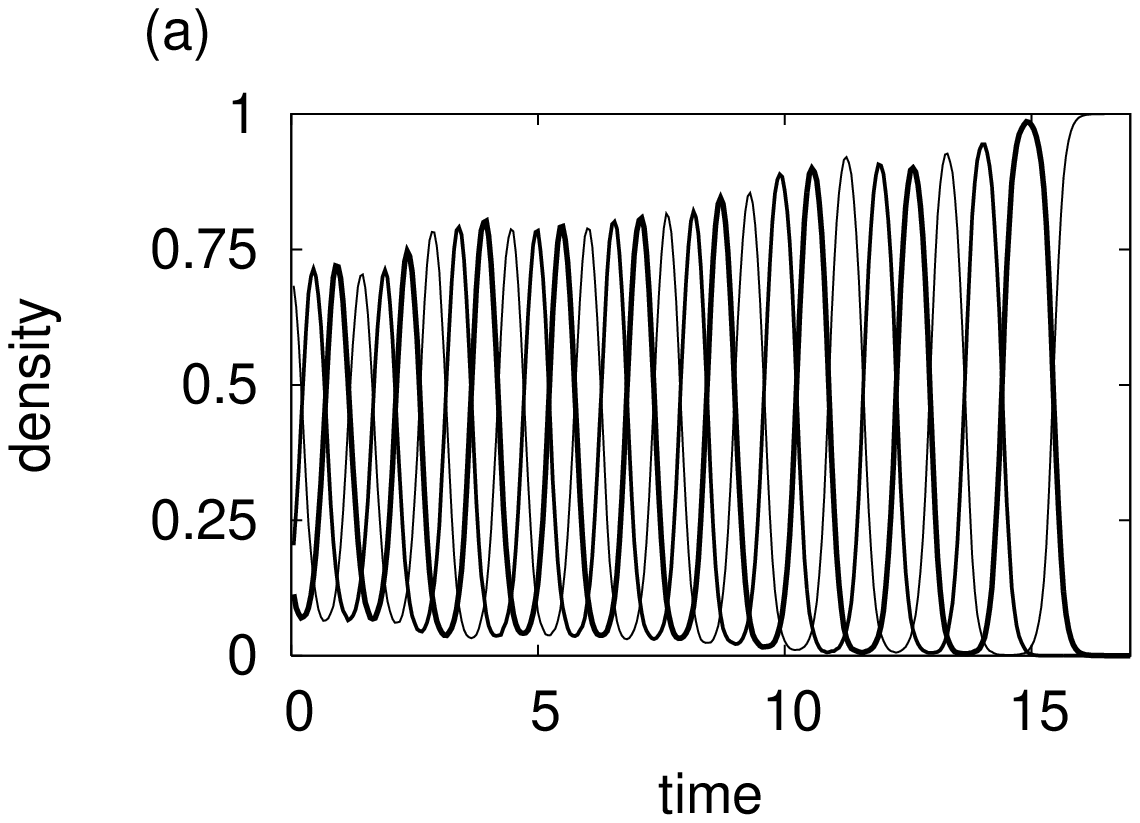}
\includegraphics[height=3in,width=3in]{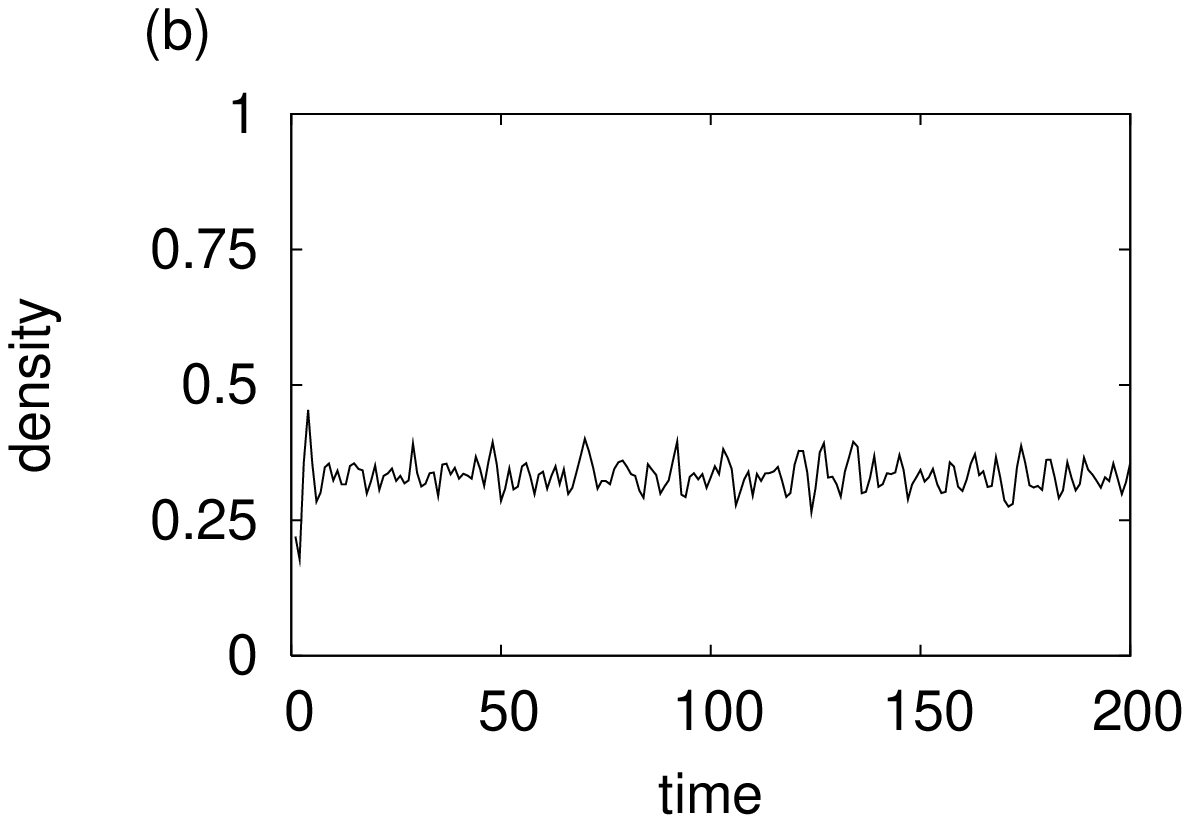}
\includegraphics[height=3in,width=3in]{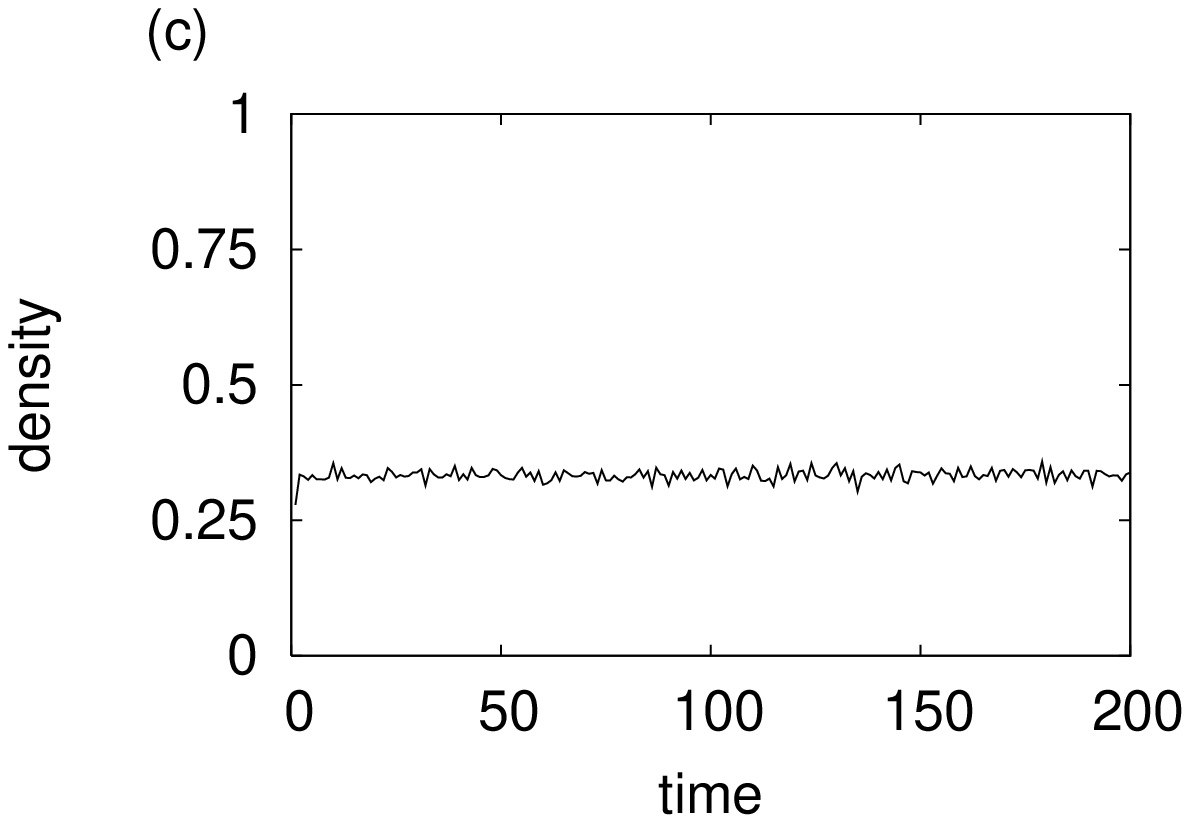}
\includegraphics[height=3in,width=3in]{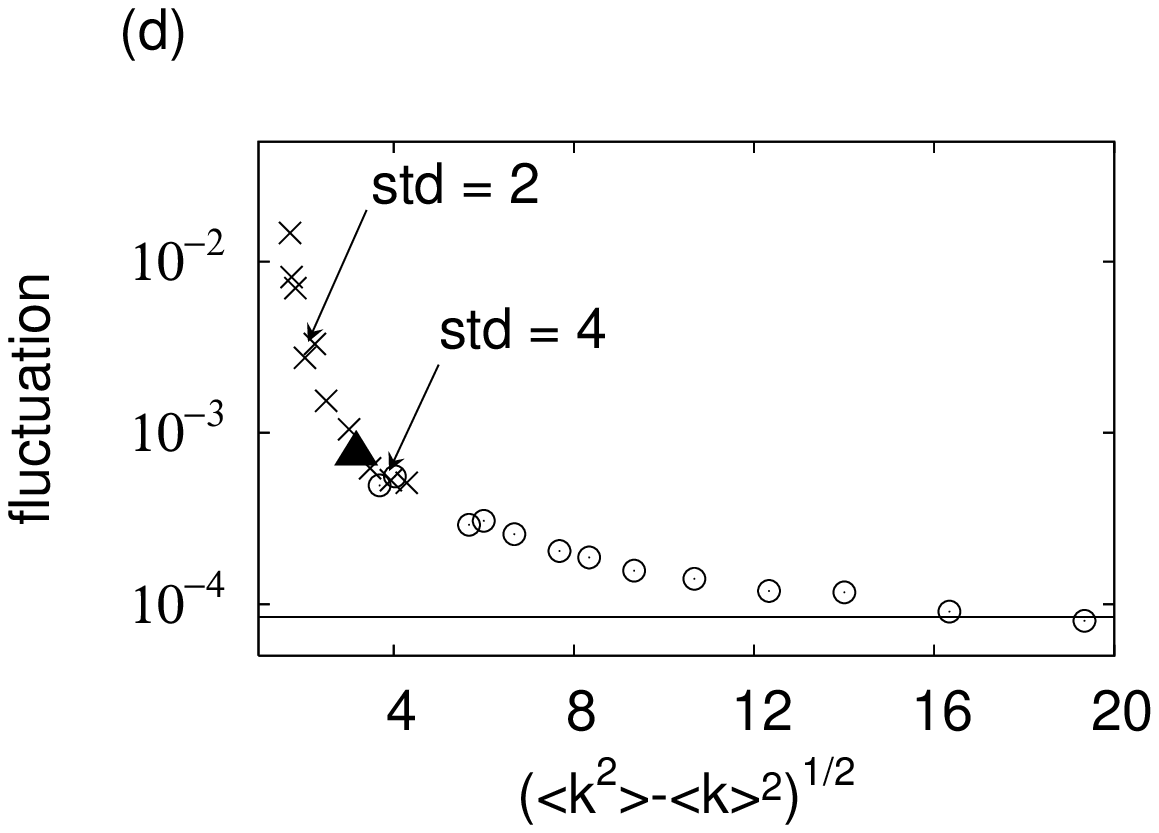}
\caption{}
\label{fig:rsp}
\end{center}
\end{figure}

\begin{figure}
\begin{center}
\includegraphics[height=3in,width=3in]{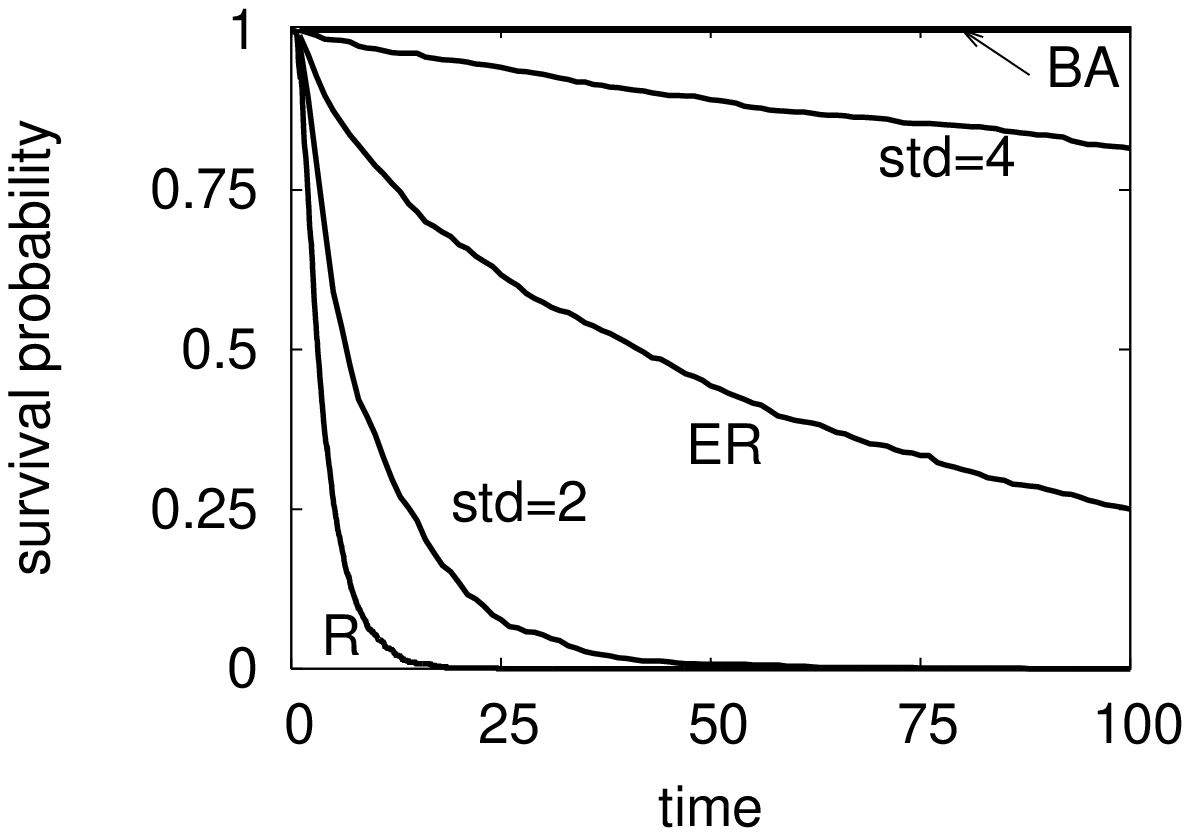}
\caption{}
\label{fig:rsp_surv}
\end{center}
\end{figure}

\begin{figure}
\begin{center}
\includegraphics[height=3.25in,width=3.25in]{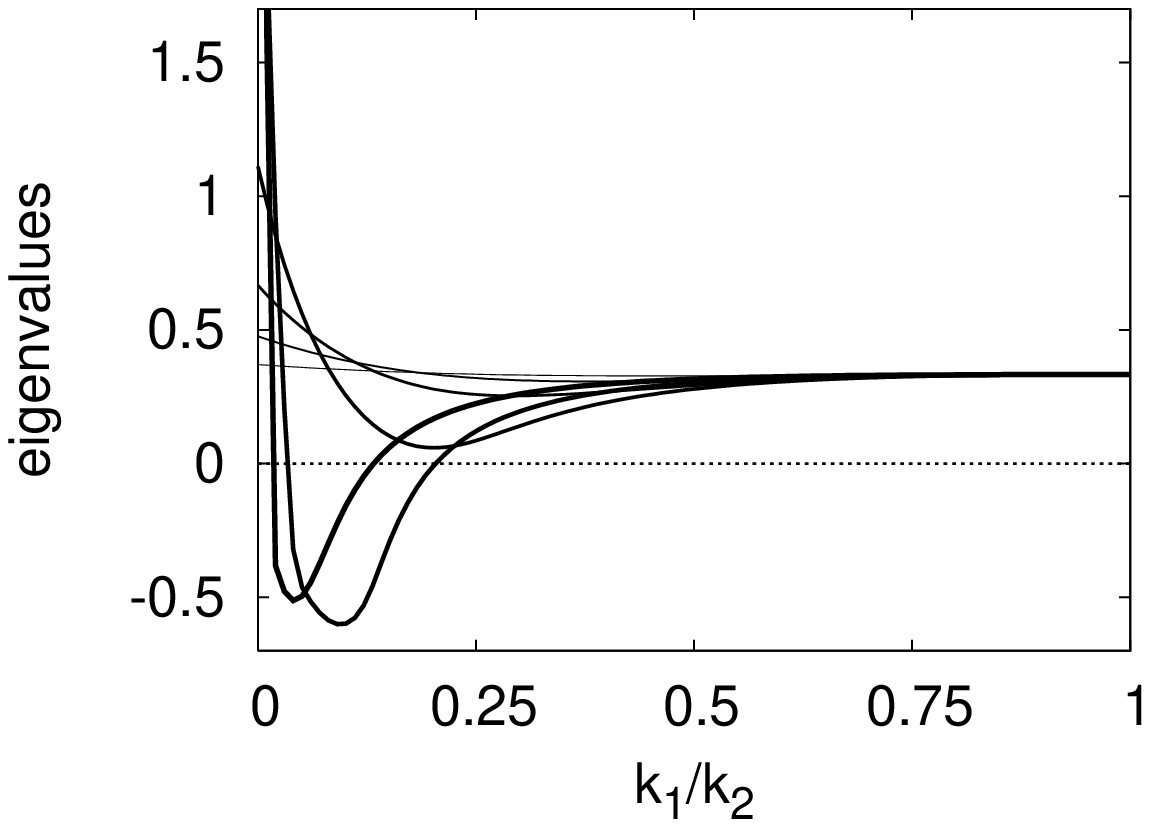}
\caption{}
\label{fig:eigs_may}
\end{center}
\end{figure}

\begin{figure}
\begin{center}
\includegraphics[height=3in,width=3in]{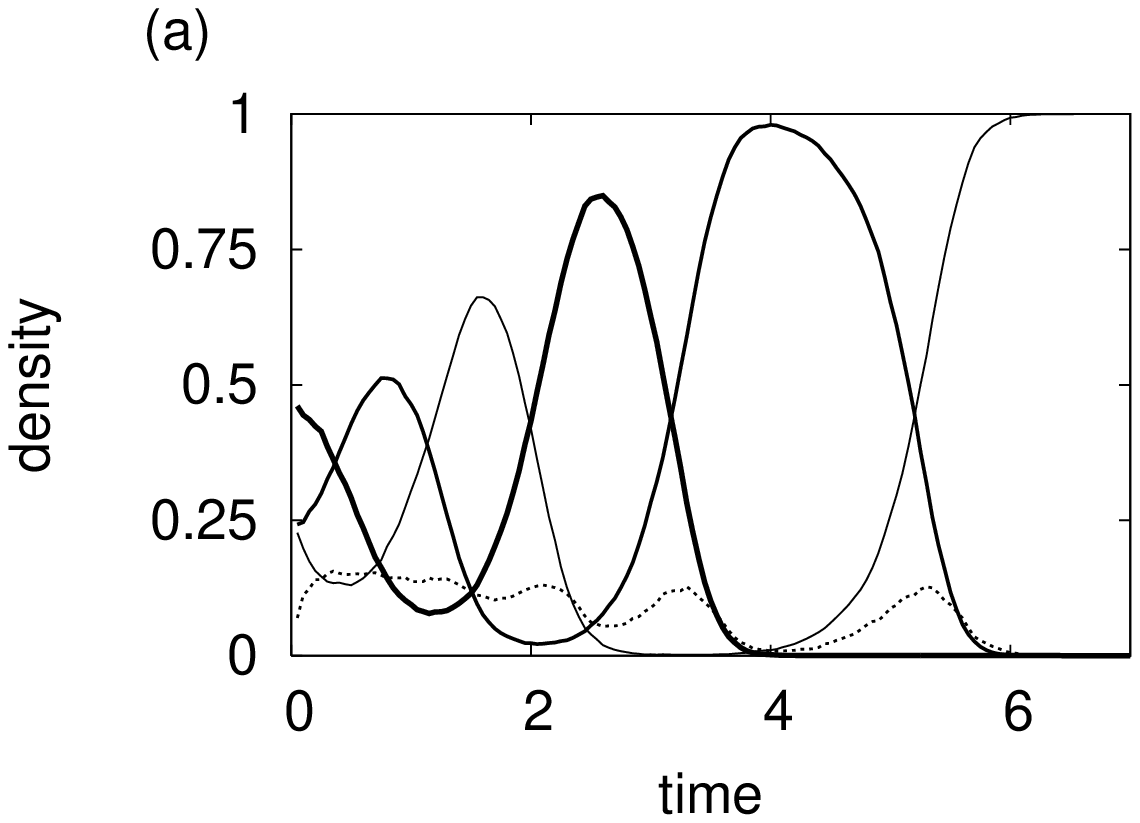}
\includegraphics[height=3in,width=3in]{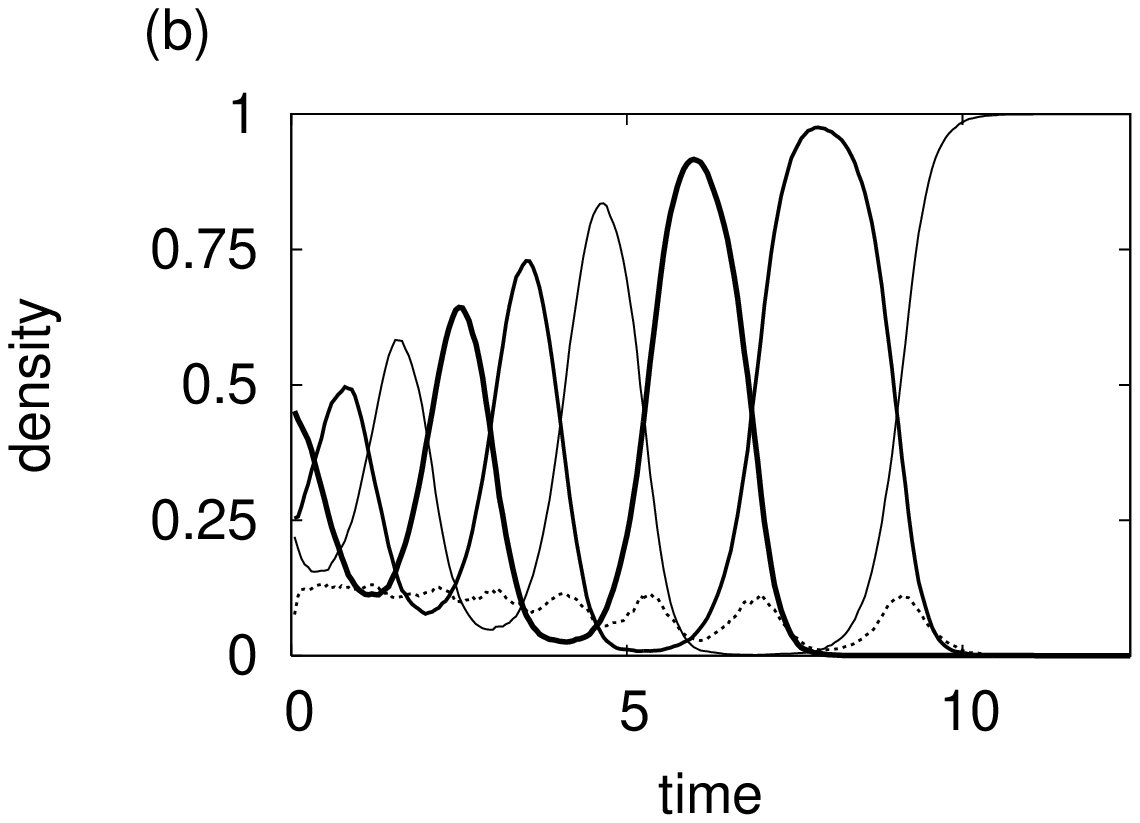}
\includegraphics[height=3in,width=3in]{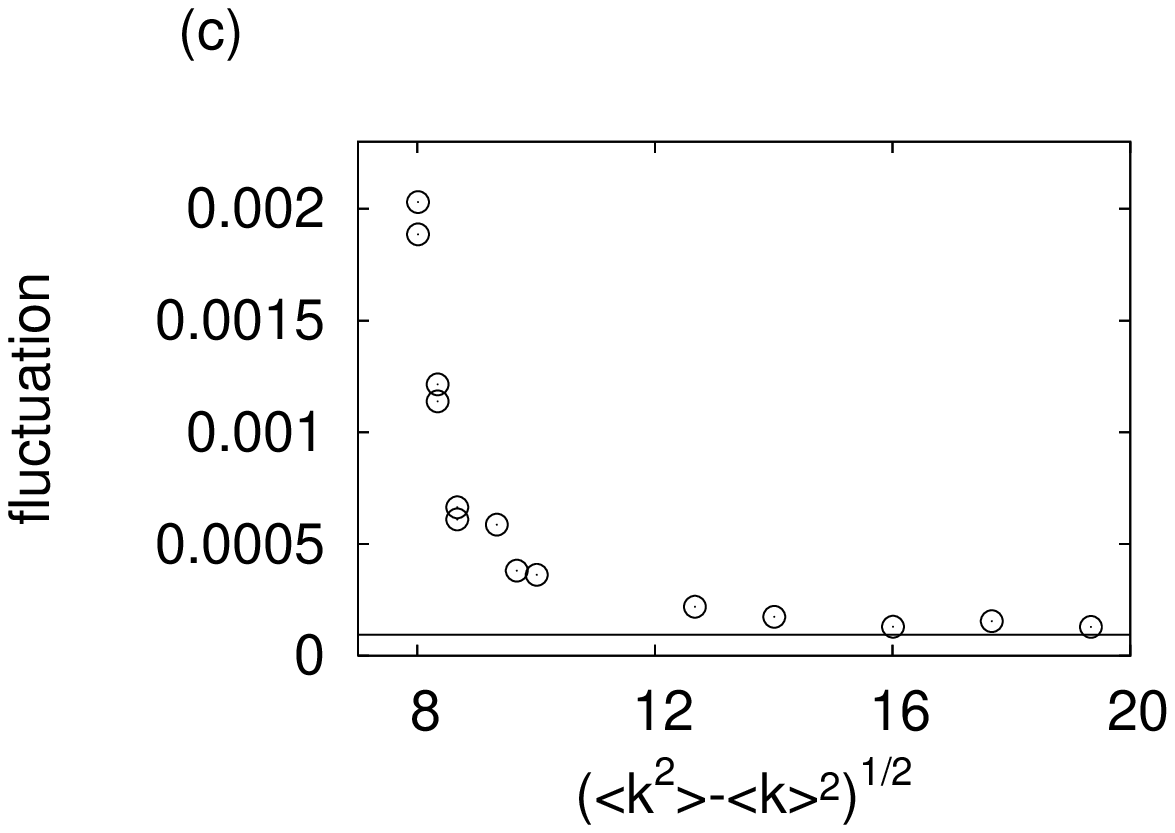}
\caption{}
\label{fig:may}
\end{center}
\end{figure}

Figure 2: Survival probabilities for 
the R random graph, the ER random
graph, the BA model, and the networks with 
Gaussian $p_k$ with standard deviation 2 and 4 [corresponding to
the crosses marked by arrows in \FIG\ref{fig:rsp}(d)]. We set
$n=200$ and $\left<k\right>=10$ for all the networks.
The survival probabilities are
calculated based on 1000 runs.

\bigskip

Figure 3: Stability of the coexistence solution of the ML model.
Real parts of the largest eigenvalues
of the Jacobian matrix obtained from \EQ(\ref{eq:eig_may}) 
with $\alpha = 2/3$ and $\beta = 2$
are presented
for $p=0.1$ (thinnest line), $p=0.3$, $p=0.5$,
$p=0.7$, $p=0.9$, and $p=0.95$ (thickest line).
We set $\left<k\right>=1$ for normalization.

\bigskip

Figure 4: ML dynamics on networks with $n=5000$, $\left<k\right>=10$,
$\alpha = 2/3$, and $\beta=2$.  The initial condition is given by
$\rho_0 = 0$, $\rho_1 = \rho_2 = 0.25$, and $\rho_3 = 0.5$.
The R random graph (a) and the ER random graph (b) do not allow
stable coexistence ($\rho_0$, dotted lines; $\rho_1$, thin solid
lines; $\rho_2$, moderate solid lines; $\rho_3$, thick solid lines).
(c) Fluctuation of population density (BA, horizontal line
[$\left(\left<k^2\right>-\left<k\right>^2\right)^{1/2} = 157.8$];
two-point $p_k$,
circles), defined by the variance of $\rho_i$ from time 150 through
300 averaged over $i=1, 2$, and $3$.

\newpage
\clearpage

\clearpage

\clearpage

\clearpage

\end{document}